\documentclass[conference]{IEEEtran}
\IEEEoverridecommandlockouts
 
\usepackage{cite}
\usepackage{amsmath,amssymb,amsfonts}
\usepackage{algorithmic}
\usepackage[ruled,norelsize]{algorithm2e}
\usepackage{graphicx}
\usepackage{textcomp}
\usepackage{xcolor}
\usepackage{hyperref}
\def\BibTeX{{\rm B\kern-.05em{\sc i\kern-.025em b}\kern-.08em
    T\kern-.1667em\lower.7ex\hbox{E}\kern-.125emX}}
    
\begin{document}

\title{AntiDeepFake: AI for Deep Fake Speech Recognition}

\author{\IEEEauthorblockN{Enkhtogtokh Togootogtokh} 
\IEEEauthorblockA{\textit{Technidoo Solutions Lab} \\
\textit{Technidoo Solutions Germany and Mongolian University of Science and Technology}\\
Bavaria, Germany \\
enkhtogtokh.java@gmail.com, togootogtokh@technidoo.com} 
\and 
\IEEEauthorblockN{Christian Klasen}
\IEEEauthorblockA{\textit{Technidoo Solutions Lab} \\
\textit{Technidoo Solutions Germany }\\
Bavaria, Germany \\
klasen@technidoo.com}
}

\maketitle

\begin{abstract}
 In this research study, we propose a modern artificial intelligence (AI) approach to recognize deepfake voice, also known as generative AI cloned synthetic voice. Our proposed AI technology, called AntiDeepFake, consists of all main pipelines from data to evaluation in the whole picture. We provide experimental results and scores for all our proposed methods. The main source code for our approach is available in the provided link:~\href{https://github.com/enkhtogtokh/antideepfake}{https://github.com/enkhtogtokh/antideepfake} repository. 
\end{abstract}


\section{Introduction}\label{intro}
Deepfake technology refers to the use of artificial intelligence and machine learning algorithms to create, manipulate, or enhance digital images, videos, and audio recordings in a way that makes them appear real and authentic. This technology has the potential to be used for a variety of purposes, including entertainment, marketing, and even political propaganda. However, it also has the potential to be used for malicious purposes, such as spreading misinformation, blackmail, and identity theft.

One of the main concerns about deepfake technology is its ability to create highly convincing fake videos and audios that can be used to manipulate public opinion and spread false information. As example, a deepfake video and audio could be used to make it appear as though a politician said something they never actually said, or to make it appear as though a celebrity endorsed a product they never actually used. This could have serious consequences for the reputation and credibility of the individuals involved, as well as for the wider public.

Another concern about deepfake technology is its potential to be used for identity theft. With the ability to create highly convincing fake audios and videos, it is possible for someone to impersonate another person and use their identity to commit crimes or access sensitive information. This could have serious legal and financial consequences for the individuals involved, as well as for the wider society.

In addition to these concerns, deepfake technology also raises important ethical questions about the use of artificial intelligence and the potential for technology to be used to manipulate and deceive people. As this technology continues to develop, it is important to consider the potential risks and benefits, and to develop appropriate safeguards and regulations to ensure that it is used in a responsible and ethical manner.

With current generative AI technology, deepfake technology has become increasingly prevalent, with the ability to create highly convincing audio and video clones of individuals. This has raised concerns about the potential for misuse and the need for effective methods to detect and prevent the spread of deepfakes. In this paper, we propose a modern artificial intelligence (AI) approach to recognize deepfake voice, which we call AntiDeepFake.

The AntiDeepFake system consists of five main pipelines, from data pipe to performance evaluation. The first pipeline involves collecting a large dataset of real and deepfake audio samples, which are then preprocessed feature extraction and engineering to data split. The second pipeline involves training a gradient boosted and tabular encoding decoding architecture to classify the audio samples as either real or deepfake.  

We present experimental results and scores for the AntiDeepFake system, which demonstrate its effectiveness in recognizing deepfake voice. Our results show that the system achieves high accuracy and robustness, even when faced with challenging edge cases and synthetic deepfake samples. We also provide the main source code for the AntiDeepFake system, which can be used by other researchers and developers to build upon our work and further improve the performance of deepfake voice recognition systems.

Deepfake technology poses a significant threat to the integrity of audio and video content, and there is a urgent need for effective methods to detect and prevent the spread of deepfakes. Our proposed AI approach,AntiDeepFake, provides a robust and accurate solution for recognizing deepfake voice, and has the potential to significantly improve the security and reliability of digital media.

Concretely, the key contributions of the proposed work are:

\begin{itemize}
\item The industry level AI technology for AntiDeepFake
\item The tabular AI framework for tabular classification, regression, and other tasks
\end{itemize}

Systematic experiments conducted on real-world acquired data have shown as: 
\begin{itemize}
\item It is possible to be common framework for many type of deep fake speeches and synthetic data recognition.
\item It is possible to achieve 99.9\% accuracy on well prepared training data to recognize.
\end{itemize}

The rest of the paper is organized as follows.
The proposed framework is described in Section \ref{proposedarch}.
The data pipeline is explained in Section \ref{datapipeline}. 
The state-of-the-art (SOTA) models are explained in Section \ref{model}. 
The details about the experimental results and evaluations are presented in Section \ref{experimentalresult}.
Finally, Section \ref{conclusion} provides the conclusions and future work.

\begin{figure*}

 \center

  \includegraphics[width=\textwidth]{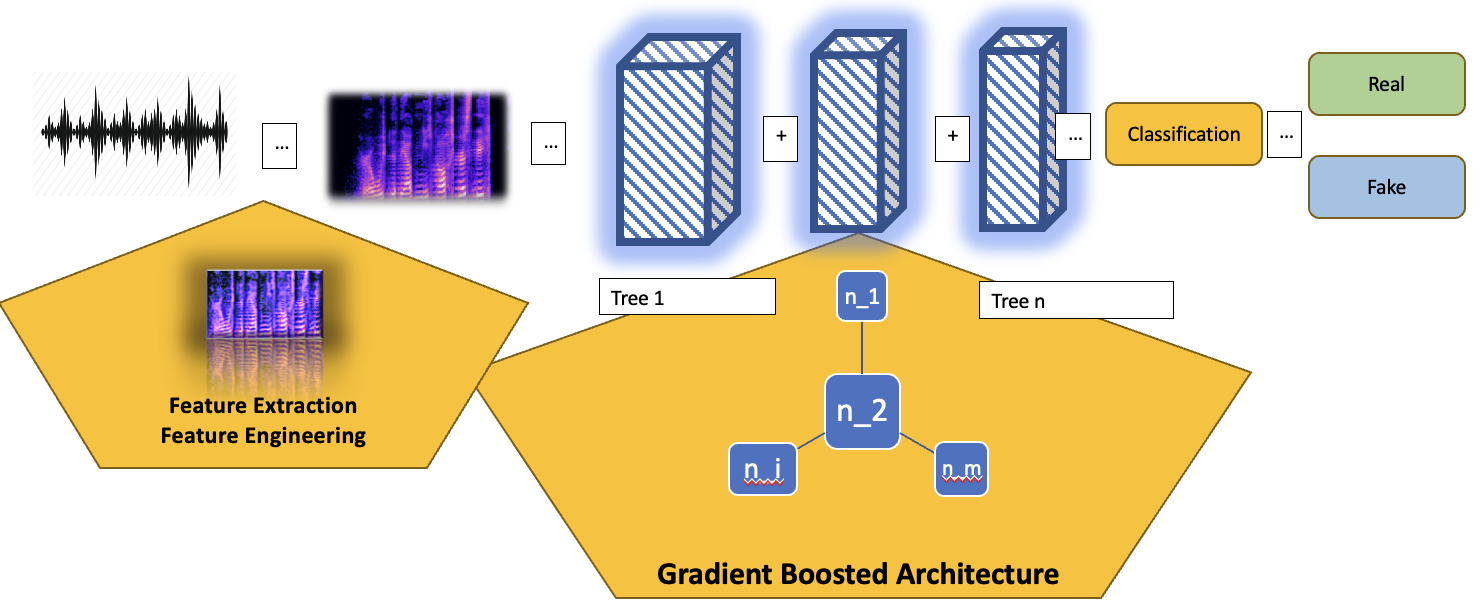}

  \caption{The proposed AntiDeepFake architecture (AntiDeepFake)}

  \label{fig:framework}

\end{figure*}

\section{The proposed Anti DeepFake Voice Architecture (AntiDeepFake)}\label{proposedarch}
 
\begin{figure}[!hb]
  \centering
  \includegraphics[width=\linewidth]{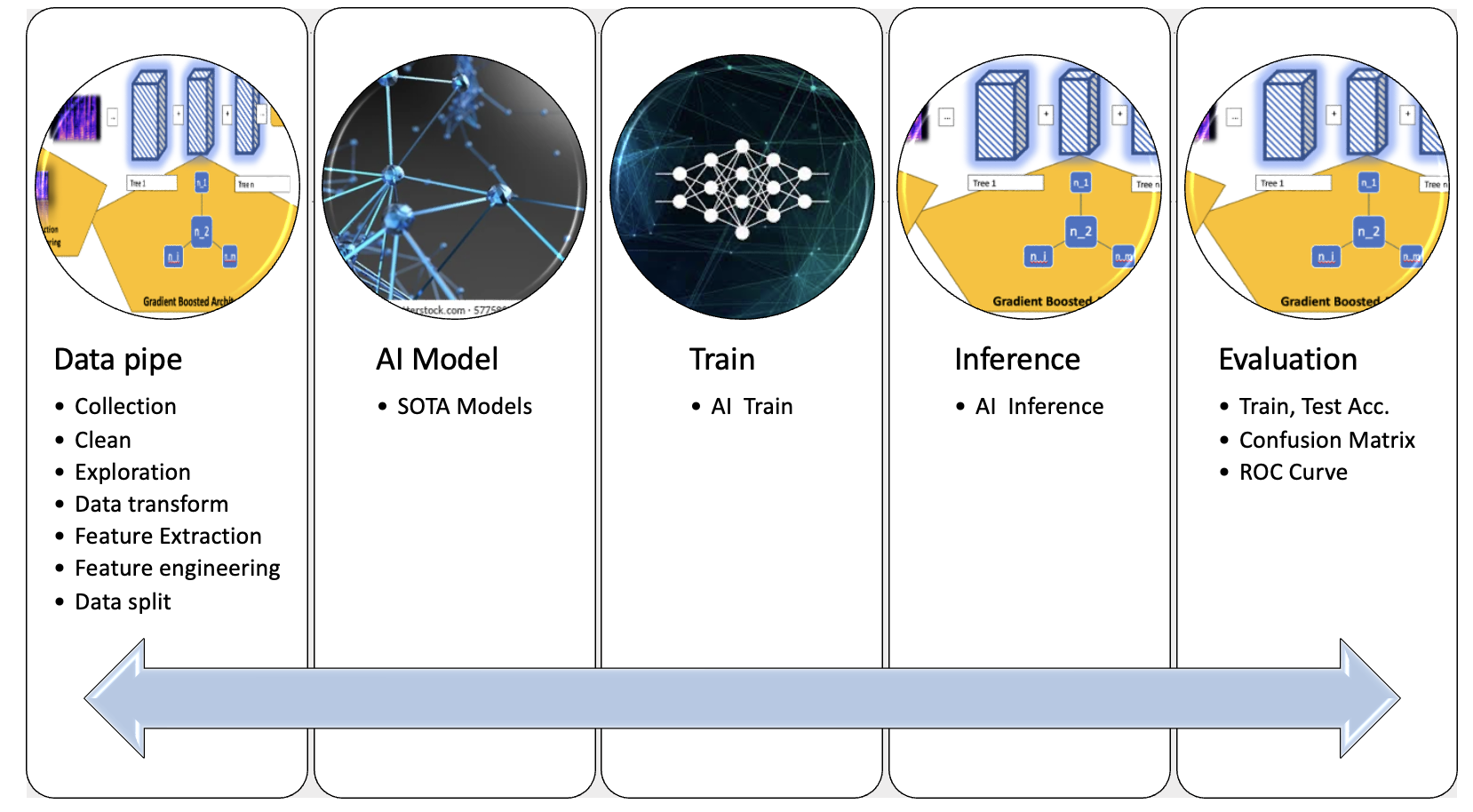}
  \caption{The Main AI Pipelines}
 
  \label{fig:pipes}
\end{figure}

In this section, we discuss the proposed AntiDeepFake approach for deepfake speech recognition as shown in Figure \ref{fig:framework}.
The AntiDeepFake has five main pipelines which are the data processing to extract efficient features and AI evaluation and experimental results  as shown in Figure \ref{fig:pipes}. We discuss them in detail with coming sections.

\subsection{The data pipeline}\label{datapipeline}
Audio significant feature extraction is the important part of modern deep learning. There are many mechanisms to do it. Here we extract melspectrogram audio feature later to train machine with high accuracy.  
\subsubsection{The data collection}\label{datacollection}
To collect an audio dataset labeled as genuine simply as real and spoof simply as fake, several methods can be employed, including crawling technologies and existing audio resources with ethical considerations. 
\begin{itemize}
\item 1. For real audio datasets, speech datasets such as ljspeech and public speeches can be utilized, as they are readily available and have no restrictions, particularly for academic research purposes. 
\item 2. On the other hand, collecting fake audio dataset is relatively straightforward due to the advancements in modern synthetic voice cloning AI models, such as recent state-of-the-art generative synthetic voice cloning AI with generative pretrained transformer (GPT) architecture technologies.
\end{itemize}
\subsubsection{The data transformation}\label{datatransformation}
Furthermore, the audio voice feature extraction process will result in tabular data, which we will discuss in detail in the following sections.
\subsection{The Feature Extraction}\label{featureextraction}
 In order to analyze the unique characteristics of real human voice, it is necessary to extract all relevant features from an audio dataset. Audio significant feature extraction is the important part of modern AI and deep learning \cite{enkhtogtokh}. In this study, we have extracted the most significant features from the dataset for the purpose of further feature engineering. The extracted features, as presented in Table \ref{tab:featureextraction}, will be utilized in subsequent analyses.   

\begin{table} [!ht]
\centering  
\begin{tabular}{||c c||} 
 \hline
 Feature Name & Variations \\  
 \hline\hline
 Pitch & STD, Mean \\ 
 \hline
 Shimmer & STD, Mean \\
 \hline
 Jitter & STD, Mean  \\
 \hline
 Formants & F1, F2, F3, F4\\
 \hline
 Chroma STFT &STD, Mean   \\
 \hline
 RMSS &STD, Mean   \\
 \hline
 Spectral Centroids &STD, Mean   \\
 \hline
 Spectral Bandwidths &STD, Mean   \\
 \hline
 Rolloffs &STD, Mean  \\
 \hline
 Zero Crossing Rates &STD, Mean   \\
 \hline
 MFCC &STD, Mean   \\
 \hline
\end{tabular}
\caption{The feature extraction from audio dataset. STD - Standard deviation}
\label{tab:featureextraction}
\end{table}







\subsection{The Feature Engineering (Selection)}\label{featureengineering}
In practice, it is not necessary to consider all extracted features, rather, it is more appropriate to focus on the most significant features, which can be referred to as feature selection or engineering.
The importance of feature engineering lies in the fact that the quality and relevance of the features used in a model can greatly affect its ability to accurately predict outcomes. In addition, feature engineering can help to address issues such as missing data and outliers, which can be common in real-world datasets. By handling these issues, the model can be more robust and better able to handle variations in the data.

Overall, feature engineering plays a critical role in machine learning by helping to ensure that the model is able to accurately capture the underlying patterns and relationships in the data, and by improving the efficiency and accuracy of the model.

In general, there are three primary categories of methodology for feature engineering:
\begin{itemize}
\item Filter. As example, Pearson
\item Wrapper. As example, Custom Boosted Tree based new approach (RFE) – SOTA
\item Embedded. As example,  Lasso Regularization
\item Ensemble. As example,  Combination of above 3 methodolies
\end{itemize}

In this study, we applied a novel custom wrapper category methodology as a gradient boosted Recursive feature elimination (RFE) approach. In our forthcoming research paper, we will elaborate on this custom feature engineering methodology.

The success of any AI model in addressing a given problem is highly dependent on the quality and quantity of the data pipeline used. Therefore, it is crucial to have a well-proposed data pipeline in place before implementing any SOTA AI models or other advanced techniques.

The following is the algorithm for obtaining  data processing:
\begin{algorithm}
\caption{Data\textunderscore Process // AntiDeepFake Data Process //} \label{alg:dataprocess}
\begin{algorithmic}[1]
 
\STATE data = collect\textunderscore data() // Collect data
\STATE data = clean\textunderscore data(data) // Clean data
\STATE explore\textunderscore data(data) // Explore data
\STATE data = extract\textunderscore features(data) // Data Transformation and Extract features
\STATE data = select\textunderscore features(data)
\STATE X\textunderscore train,y\textunderscore train,X\textunderscore test,y\textunderscore test = train\textunderscore test\textunderscore split(data)
\STATE return X\textunderscore train,y\textunderscore train,X\textunderscore test,y\textunderscore test
\texttt{\\}
\texttt{\\}
\end{algorithmic}
\end{algorithm}

\subsection{The AI modeling}\label{model}  
In the field of tabular AI classification, gradient boosted AI models and deep tabular data learning architectures that incorporate sequential attention models have emerged as the dominant approaches. Which means state-of-the-art models are as example CatBoost\cite{catboost} TabNet\cite{tabnet} , XGBoost\cite{xgboost} . The expermintal results discuss in Section \ref{experimentalresult}. It will show how their comparision and evaluation results.

\subsection{The AI training}\label{training}  
In our research, we train on all state-of-the-art (SOTA) models described in Section \ref{model}, and the main training algorithm employed is presented below:
\begin{algorithm}
\caption{Train  // AntiDeepFake Train //} \label{alg:train}
\begin{algorithmic}[1] 
\STATE X\textunderscore train, y\textunderscore train, X\textunderscore test, y\textunderscore test = Process\textunderscore Data()
\STATE sota\textunderscore models = [TabNet(),XGBoost(),CatBoost()]
\STATE model = sota\textunderscore models[i], i= selected index
\texttt{\\}
\STATE  model.fit(X\textunderscore train,y\textunderscore train)
\STATE  model.save(model\textunderscore save\textunderscore path)
\end{algorithmic}
\end{algorithm}

\subsection{The inference}\label{inference}
After training process,  the main inference algorithm employed is presented below:
\begin{algorithm}
\caption{Inference  // AntiDeepFake Inference //} \label{alg:inference}
\begin{algorithmic}[1] 
\STATE model = load(model\textunderscore save\textunderscore path)
\STATE y\textunderscore hat = model.predict(inference\textunderscore data)
\STATE return y\textunderscore hat
\texttt{\\}
\end{algorithmic}
\end{algorithm} 

\newpage
\section{Experimental Results and Evaluation}\label{experimentalresult}
In this section, we present the setup of our research and then evaluate the state-of-the-art (SOTA) models using the selected features. We conduct experiments in systematic scenarios to analyze the performance of the models.
\subsection{Setup}
We train and test on ubuntu 18 machine with capacity of (CPU: Intel(R) Xeon(R) CPU @ 2.20GHz, RAM:16GB, GPU: NVidia GeForce GTX 1070, 16 GB). 

\subsection{The dataset}
In this research paper, we present a dataset that was collected in accordance with the description provided in Section \ref{datacollection}. The dataset contains two distinct speech labels, namely real and fake speeches.

\subsection{The AI Scores}
Along with classification accuracy, further metrics are also considered for model comparison. These are precision, which measures the rate at which predicted positives are correct among all positive predictions, and allows for false-positive analysis.
Precision is important since a high precision would minimise false accusations of AI-generated speech when the audio is, in fact, natural voice. Similarly, recall which is a measure of how many positive cases are correctly predicted, which enables analysis of false-negative predictions.
Higher recall suggests that the model is not falsely classifying AI-generated speech as human speech. These results are then combined to compute the F-1 score.

The accuracy of training and testing for SOTA models is presented in Table \ref{tab:recacc}. Among the models evaluated, the CatBoost model achieved the highest score

\begin{table} [!ht]
\centering  
\begin{tabular}{||c c c||} 
 \hline
 AI model & Training Accuracy & Testing Accuracy \\  
 \hline\hline
 CatBoost\cite{catboost} & 1.0 & 0.937  \\ 
 \hline
 XGBoost\cite{xgboost} & 1.0 & 0.925  \\ 
 \hline
 TabNet\cite{xgboost} & 0.86 & 0.80  \\ 

 \hline
\end{tabular}
\caption{The SOTA model training and testing accuracy.}
\label{tab:recacc}
\end{table}

The following table \ref{tab:reccatboost} presents the main scores for the CatBoost\cite{catboost} model.
\begin{table} [!ht]
\centering  
\begin{tabular}{||c c c c||} 
 \hline
 Name & Precision & Recall & F1-score \\  
\hline\hline
Real   &    0.94   &   0.93   &   0.94\\  
\hline
Fake    &    0.93  &     0.94  &     0.94 \\ 
\hline
Accuracy  &     &       &         0.94 \\ 
\hline
Macro avg. &  0.94  &     0.94  &     0.94 \\ 
\hline
Weighted avg.   &  0.94  &     0.94  &     0.94 \\
\hline
\hline
\end{tabular}
\caption{The CatBoost model confusion scores.}
\label{tab:reccatboost}
\end{table}
\begin{figure}[!ht]
  \centering
  \includegraphics[width=\linewidth]{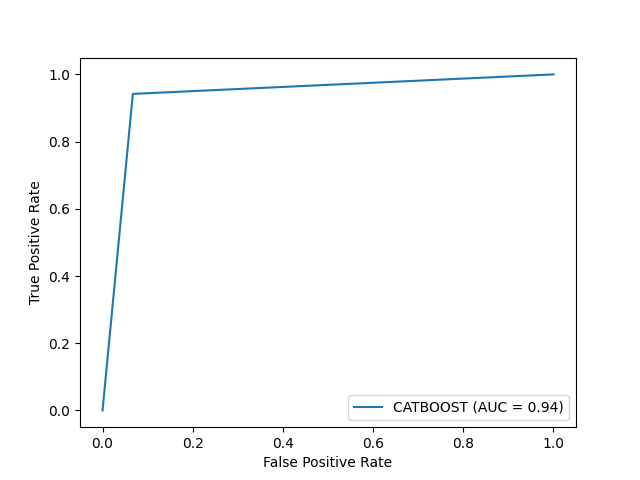}
  \caption{The ROC Curve of CatBoost} 
  \label{fig:catboost_roc}
\end{figure}

The following table \ref{tab:recxgboost} presents the main scores for the XGBoost\cite{xgboost} model.

\begin{table} [!ht]
\centering  
\begin{tabular}{||c c c c||} 
 \hline
 Name & Precision & Recall & F1-score \\  
\hline\hline
Real &  0.94 &     0.91   &   0.92 \\  
\hline
Fake &  0.91  &    0.94    &  0.93 \\ 
\hline
Accuracy  &     &       &        0.93 \\ 
\hline
Macro avg.& 0.93   &   0.93    &  0.93 \\ 
\hline
Weighted avg. & 0.93   &   0.93    &  0.93 \\
\hline
\hline
\end{tabular}
\caption{The XGBoost model confusion scores.}
\label{tab:recxgboost}
\end{table}

\begin{figure}[!ht]
  \centering
  \includegraphics[width=\linewidth]{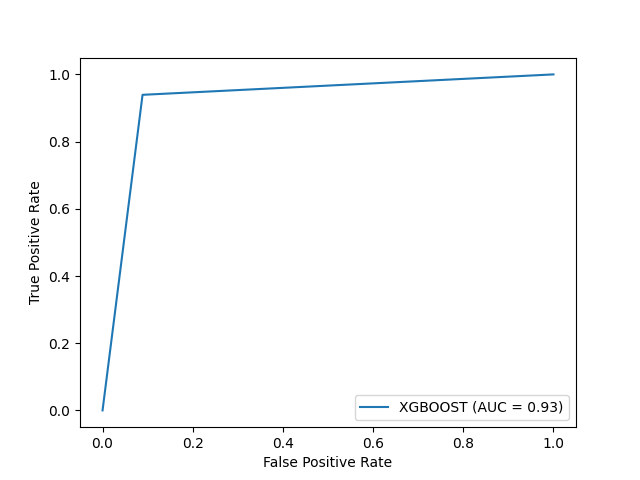}
  \caption{The ROC Curve of XGBoost} 
  \label{fig:xgboost_roc}
\end{figure}

The following table \ref{tab:rectabnet} presents the main scores for the TabNet\cite{tabnet} model.
\begin{table} [!ht]
\centering  
\begin{tabular}{||c c c c||} 
 \hline
 Name & Precision & Recall & F1-score \\  
\hline\hline
Real   &    0.83   &   0.78  &    0.80 \\  
\hline
Fake   &    0.79  &    0.84  &    0.81 \\ 
\hline
Accuracy  &     &       &         0.81\\ 
\hline
Macro avg.  &      0.81  &     0.81    &   0.81 \\ 
\hline
Weighted avg.   &    0.81   &   0.81  &    0.81 \\
\hline
\hline
\end{tabular}
\caption{The TabNet model confusion scores.}
\label{tab:rectabnet}
\end{table}
 
\begin{figure}[!ht]
  \centering
  \includegraphics[width=\linewidth]{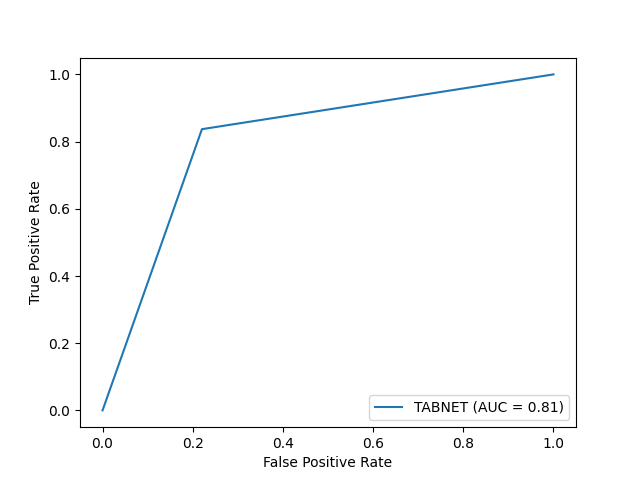}
  \caption{The ROC Curve of TabNet}
 
  \label{fig:tabnet_roc}
\end{figure}

\newpage
\section{Conclusion}\label{conclusion}
This research paper focuses on the recognition of generative AI, specifically, those related to deepfake with AI-generated synthetic human speech. The contributions of this study include the development a comprehensive analysis of the modern AI pipeline significance of audio features extraction and engineering from real and AI-generated synthetic speech, and the SOTA of AI models that can predict the both real and synthetic of speech.
In future works, we will publish next series of research to apply on AntiDeepFake for video analysis tasks.

\vspace{12pt}

\end{document}